\documentclass[aps,pre,twocolumn,superscriptaddress]{revtex4-1}

\usepackage{amsmath}
\usepackage{graphicx}
\usepackage{hyperref}
\hypersetup{
    colorlinks=true,        % false: boxed links; true: colored links
    linkcolor=blue,         % color of internal links
    citecolor=blue,         % color of citations
    urlcolor=blue           % color of external links (URLs)
}
\usepackage{dcolumn}
\usepackage{subfigure}
\usepackage{algorithm}
\usepackage{algpseudocode}
\usepackage[english]{babel}
\usepackage[autostyle, english = american]{csquotes}
\MakeOuterQuote{"}

\begin{document}

\title{Evidence of Non-Equilibrium Critical Phenomena in a Simple Model of Traffic} %

\author{Aryaman Jha}
\affiliation{Center for Nonlinear Sciences, School of Physics, Georgia Tech}
\author{Kurt Wiesenfeld}
\affiliation{Center for Nonlinear Sciences, School of Physics, Georgia Tech}
\author{Garyoung Lee}
\affiliation{School of Civil and Environmental Engineering, Georgia Tech}
\author{Jorge Laval}
\affiliation{School of Civil and Environmental Engineering, Georgia Tech}

\date{\today}

\begin{abstract}
We present a novel approach to understand vehicular traffic jams by studying a simple model, Elementary Cellular Automaton Rule 184 (ECA 184). Using key traffic observables, such as the total delay and relaxation time, as well as microscopic measures like delays and lifetimes of individual jams, we show how these quantities can fully characterize the system’s behavior, revealing features analogous to those of a continuous phase transition. We exploit specific properties of ECA 184 to develop an efficient algorithm for calculating these observables numerically and introduce an auxiliary quantity, termed "elementary jams", which allows us to determine these observables.  We discuss the implications of our results, highlighting connections to a potential field-theoretic description of traffic and suggest future application of these methods to more complex models.
\end{abstract}

\maketitle

%\input{introduction}
%----------------------------------
\section{Introduction}

Experimental research in traffic flow has revealed power laws and universal behavior in the data \cite{schrank2019urban, bettencourt2007growth, geroliminis2008existence}. Understanding these is useful for managing traffic jams and reducing congestion. To study this behavior, discrete cellular automata (CA) models provide a simple description of traffic, especially where continuum models fail to explain observations \cite{nagel1992cellular, biham1992self, chowdhury2000statistical}.

One well-known CA model is the Nagel-Schreckenberg (NS) model \cite{nagel1992cellular}, where vehicles accelerate to a maximum velocity \(v_{\text{max}}\) and decelerate randomly with probability \(p\). Analytical approaches to study the NS model rely on approximation techniques such as mean-field theory and related simplification schemes \cite{schadschneider1993cellular, vilar1994cellular, teoh2018renormalization}. Computational efforts have explored and characterized these models across various limits, focusing primarily on how steady-state properties change with the control parameter, density $\rho$ \cite{nagel1993deterministic, sasvari1997cellular, souza2009traffic, balouchi2016finite}. However, traffic jams and delays are better understood as fluctuations arising from spatiotemporal correlations and as transient effects in the deterministic limit of the traffic model, which are often missed or truncated by both analytical and numerical methods.

Nagel and Paczuski \cite{nagel1995emergent,paczuskitextordfeminine1996self} analyzed how traffic jams may form spontaneously and obtained statistics for the distribution of traffic jam lifetimes using an infinite-jam initial condition, applicable to specific scenarios. They provided heuristics for how an order parameter and response function might depend on the control parameter. Additionally, their analysis on large but finite systems limited conclusions about the system’s behavior in the thermodynamic limit.

In this paper, we build on this picture by studying the simpler model Elementary Cellular Automata Rule 184 (ECA 184) which corresponds to the limit $v_{max} = 1$ and $p = 0$ of the NS model.  We provide strong numerical evidence that the jams, which represent transients in this model, exhibit all the hallmarks of critical phenomena when analyzed appropriately.

We begin in Section II by providing background on ECA 184 and introducing the key quantities of interest through space-time plots. A new quantity, termed ``elementary jams" is defined to characterize the system's dynamics, enabling the definition of all relevant traffic quantities. In Section III, we present our numerical analysis and perform a finite-size scaling study to estimate the critical exponents in the thermodynamic limit (system size $L \rightarrow \infty$). Finally, Section IV provides a summary of our results and discusses their implications for analyzing traffic flow models.  Additionally in the Appendix we describe an algorithm which allows us to efficiently obtain traffic observables for large system sizes.

%----------------------------------

%\input{background}
%----------------------------------
\section{Background: Spacetime plots and quantities}

\subsection{A model of traffic}

Elementary Cellular Automata (ECA) are a class of 1D deterministic cellular automata where the value of each site can only be 0 or 1, with simultaneous site update rule that depends only on the two nearest neighbors.  Under these conditions there are 256 such rules possible which have been numbered from 0 to 255. This space of systems has been studied starting with Wolfram \citep{wolfram2003new} and has been further systematically explored \citep{israeli2006coarse}.  Typically, these studies use periodic boundary conditions, which we also do in this paper.

\begin{figure}[htbp]
    \centering
    \includegraphics[width=0.95\columnwidth] {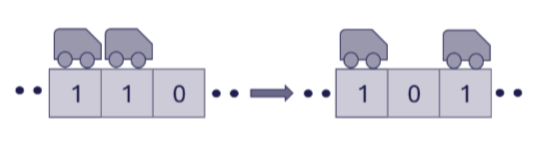} % Adjust the file name and path as needed
    \caption{Schematic showing ECA184 as a model of  drivers on a single lane.}
    \label{fig:rule}
\end{figure}

Rule 184 is a particular instance and it may also be interpreted as a one-lane traffic model: The 1D lattice may be viewed as a road where each site can either be occupied by a vehicle (\(1\)) or be empty (\(0\)). These vehicles are initialized with a density $\rho$ and they move according to the following update rule: If the right hand adjacent site is occupied, the vehicle remains stationary; otherwise, it moves forward one lattice site as illustrated in Fig.~\ref{fig:rule}.

%\subsection{Effective dynamics and traffic jams}

\begin{figure}[htbp]
    \centering
    \includegraphics[width=0.95\columnwidth]{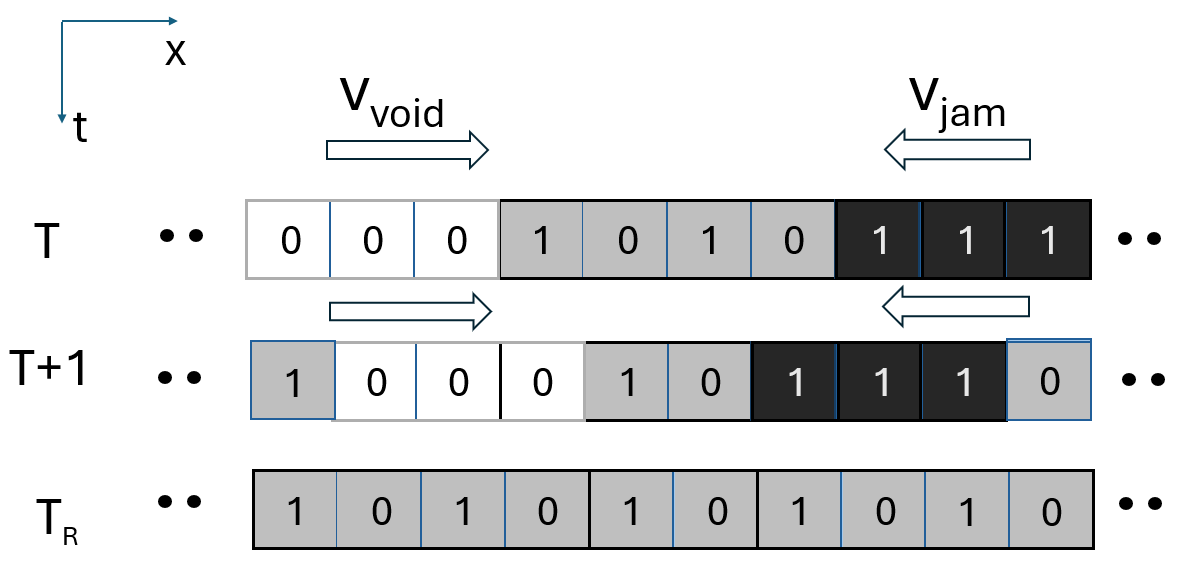} % Adjust the file name and path as needed
    \caption{The lattice at a given time step may be partitioned into jams (\textit{shown in black}) which move with a velocity of -1 units/timestep, voids (\textit{shown in white}) which move with a velocity of 1 unit/timestep as well as a capacity state (\textit{shown in grey}) which acts as a background medium.}
    \label{fig:effective dynamics}
\end{figure}

At any instant in time, the spatial configuration can be partitioned into domains of three types (see Fig.~\ref{fig:effective dynamics}): (i) continuous sequences of \(1\)s called ``jams" , (ii) continuous sequences of \(0\)s called ``voids", and (iii) alternating sequences of \(0\)s and \(1\)s called the ``capacity" state, where the flow is maximized, with vehicles propagating at maximum (unit) speed without hindrance.
We may describe an effective dynamics in terms of these jams and voids as counter-propagating particles and anti-particles which collide and annihilate, while the capacity states serve as a stable, passive background.  Eventually, either all of the particles or all of the anti-particles are destroyed, and thereafter the system simply rigidly translates at constant speed without further meaningful evolution.
For $\rho< 0.5$ the jams eventually disappear, and for values greater than $ 0.5$ we get jams that persist indefinitely moving in the direction opposite to the vehicles, while at $\rho = 0.5$ we obtain a single steady state of alternating 0s and 1s regardless of the initial condition. 

\begin{figure}[h]
    \centering
    \includegraphics[width=0.9\columnwidth]{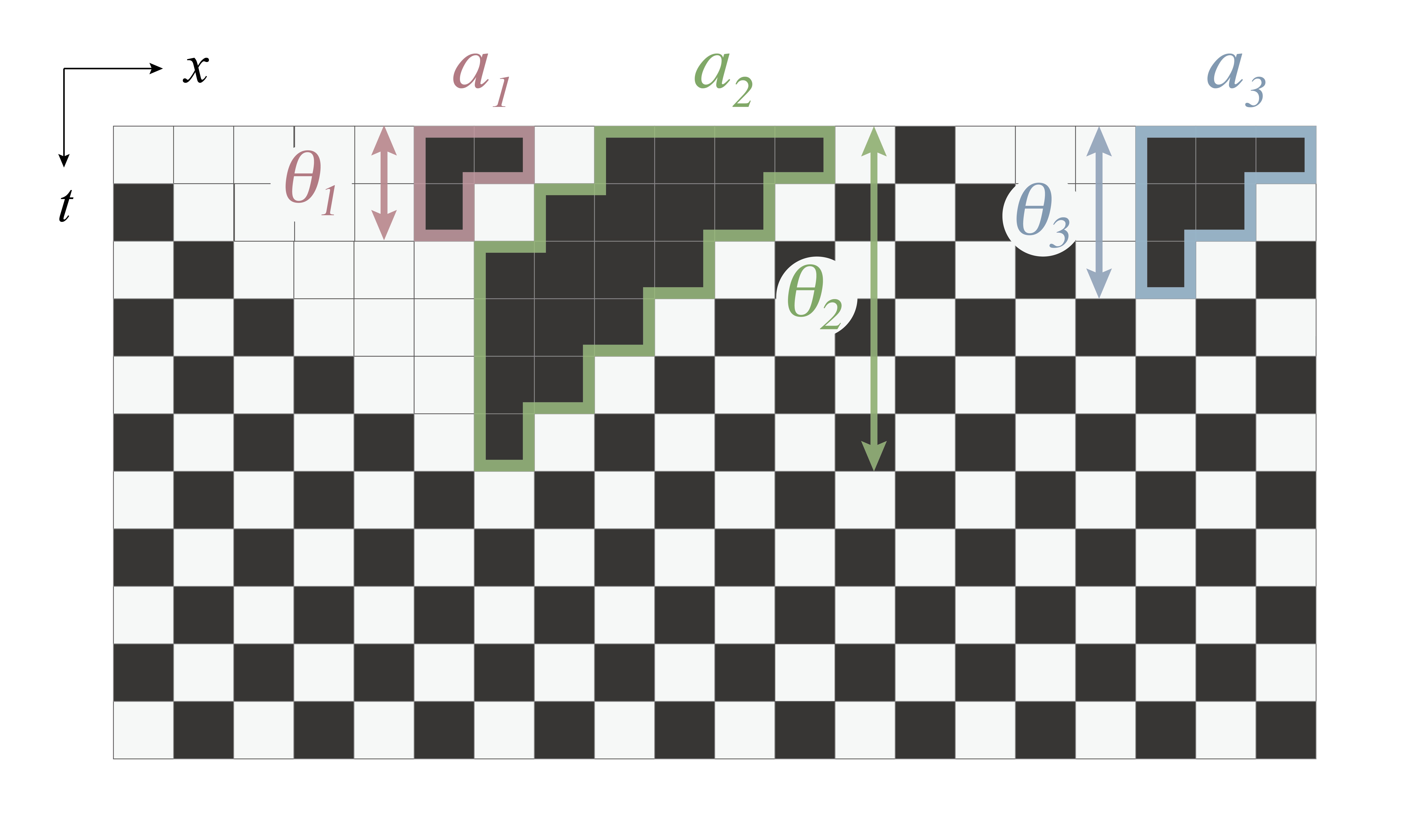}  % Adjust the file name and path as needed
    \caption{Spacetime plot of a system of size L = 20, and filling density $\rho$ = 0.5, with three clusters, with their corresponding areas $a_i$ and the height of the given cluster giving its lifetime $\theta_i$ (as indicated by the blue arrows).}
    \label{fig:example_microscopic}
\end{figure}

It is convenient to describe these dynamics via the geometric patterns generated in the space-time diagram. Regardless of initial conditions, the space-time diagrams of ECA184 have the same few simple features: voids propagate steadily in the same direction as traffic, and jams propagate in the opposite direction.  The jams form connected ``clusters" and represent an important observable analyzed here.  Finally, the checkerboard pattern arises from the propagation of capacity states.  
An example is shown in Fig. \ref{fig:example_microscopic}, for a system of length $L=20$ and filling density $\rho = 0.5$. Each horizontal row represents the spatial configuration at a fixed time; a black (occupied) site represents a vehicle; otherwise, it is an empty site. Time increases going downward. For any initial condition, the system reaches a steady state in a finite time \(T_R \leq \frac{L}{2}\).

In the context of traffic modeling, clusters are particularly significant as they represent traffic jams.  Quantitatively, each cluster is characterized by two key properties: (i) the area $a_i$, equal to the net delay suffered by the vehicles caught in the $i^{th}$ traffic jam, and (ii) the cluster height $\theta_i$, equal to the time it takes for the traffic jam to clear up. In the example shown in Fig.\ref{fig:example_microscopic}, there are three clusters, with values $(\theta_i, a_i) = (2,3), (6,18), (3,6)$ for $i = 1, 2, 3$, respectively. 

As is evident from Fig.\ref{fig:example_microscopic}, and is true for all space-time plots generated by ECA184, the cluster shapes are simple and compact. We exploited this fact to develop an efficient algorithm that allows us to determine the cluster areas and heights directly from the initial conditions, {\it i.e.} without having to explicitly compute the space-time diagram.  The algorithm, described in the Appendix, lets us analyze large system sizes with relatively modest computational effort. 

%\subsection{Space-time plots}

%add section about our interest rather in a space-time characterization of ECA184...
%Rather than probing the effective dynamics, we may probe the system in a space-time sense. We obtain the space-time plot by running an initial condition until the dynamics settles into its periodic state.

% We note that this way of understanding particle hopping models in natural to traffic flow research \cite{laval2013hamilton, laval2022traffic}.

\noindent In addition to characterizing each cluster, we define two system-wide (``macroscopic") quantities:
(i) the total traffic delay $A$, equal to the total area of all clusters combined, and (ii) the system relaxation time $T_R$, equal to the longest-lived cluster:
\begin{equation}
    A = \sum_{i\: \epsilon\: \mathcal{C}} a_i
\end{equation}
\begin{equation}
    T_R = \max_{i\: \epsilon\: \mathcal{C}} \theta_i
\end{equation}
where $\mathcal{C}$ is the set of clusters for the given initial condition. 

As laid out in the next section, the system dynamics displays a sharp transition as a function of the density $\rho$.  Note that the number of vehicles -- and therefore $\rho$ -- is strictly conserved for ECA184 with periodic boundary conditions.  The sharp transition in macroscopic behavior is accompanied by the emergence of power-law distributions for the microscopic observables.  

\subsection{Elementary jams} \label{Elementary_jams}

\begin{figure}[h]
    \centering
    \includegraphics[width=0.95\columnwidth]{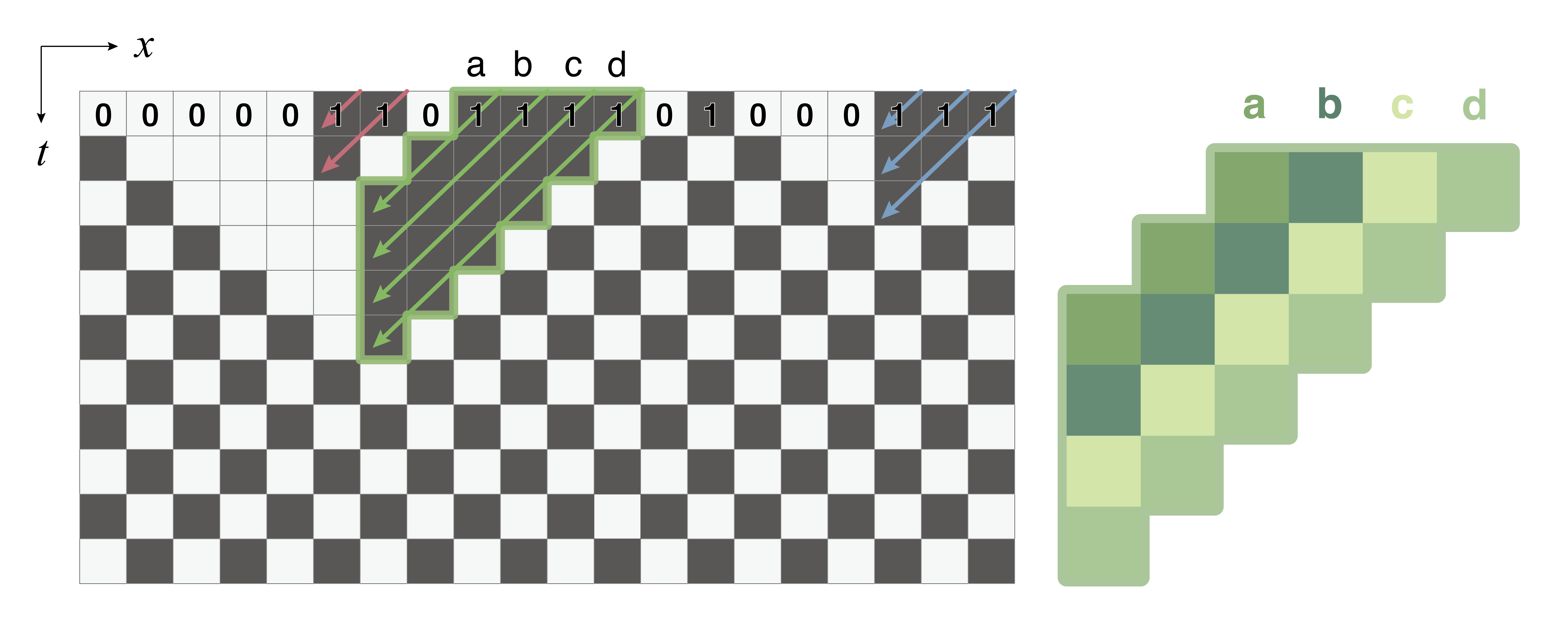} % Adjust the file name and path as needed
    \caption{(Left) Spacetime plot of a system of size L = 20, and $\rho$ = 0.5, with three clusters, and diagonal arrows showing the corresponding elementary jams $m_j$ constituting each cluster. (Right) The largest cluster is decomposed into component elementary jams of lengths $m_a, m_b, m_c, m_d$ shown as separate color components. }
    \label{fig:microjams}
\end{figure}

We now define the ``elementary jams", in terms of which we can readily compute all of the traffic flow quantities defined above \cite{laval2023self}.  First, note that all of the space-time clusters have a similar shape:  each cluster is composed of diagonally shaped elements, which we call elementary-jams, with length $m_j$ as illustrated in Fig.~\ref{fig:microjams}.

\noindent For example, the largest cluster in Fig. \ref{fig:microjams} arises from an initial jam of length 4 (sites $a, b, c, d$), and is therefore decomposed into 4 elementary-jams of lengths $\{ m_a = 3, m_b = 4, m_c = 5, m_d = 6 \}$, respectively. The cluster area is $3+4+5+6 = 18$, and the cluster lifetime is $max\{3, 4, 5, 6\} = 6$.
\noindent Note that every site in the initial condition that forms part of a jam may be associated with an elementary jam. We set $m_j$ = 0 for sites not associated with jams. 
\noindent In terms of the elementary jams, the expressions for the macroscopic quantities delay $A$ and relaxation time $T_R$ become:
\begin{equation} \label{eqn:Elementary_jams}
    A = \sum_{i = 1}^{N} m_i \:\:\:\:\: \text{and} \:\:\:\:\:  T_R = \max_{i} m_i
\end{equation}

\noindent where the index $i$ is over the lattice sites in the initial condition.  Thus, obtaining the distribution of elementary jams allows us to determine the distributions and averages of all the quantities of interest.

%----------------------------------

%\input{results}
%----------------------------------
\section{Results} \label{sec:results}
%How we get our samples/ how these results are obtained
\noindent The system consists of a one-dimensional spatial lattice of \textit{L} sites, with periodic boundary conditions. An initial condition is formed by choosing at random exactly $\rho L$ sites to be occupied. The full ensemble of $\binom{\textit{L}}{\textit{$\rho L$}}$ initial conditions is sampled \textit{X} times. For each initial condition, we obtain the set of elementary-jams from which the cluster properties are determined. The data are aggregated over the \textit{X} realizations. The entire process is repeated for different values of the density $\rho$. We find that using system sizes of $L\approx 10000$ over $X\approx 1000$ realizations provides robust sampling for our numerical analysis.
\subsection{Macroscopic observables}
%Mention the macroscopic quantities and their result
\noindent We consider first the system-wide (macroscopic) quantities, namely the total delay $\langle$A$\rangle$ and relaxation time $\langle T_R \rangle$, where the angular brackets denote the ensemble average. We normalize the total delay $\langle$A$\rangle$ by the maximum area of the space-time plot $\frac{1}{2}L^{2}$ to get the normalized delay $\phi = 2\langle$A$\rangle/L^{2}$

\begin{figure}[h! tbp]
    \centering
    \includegraphics[width=0.5\textwidth]{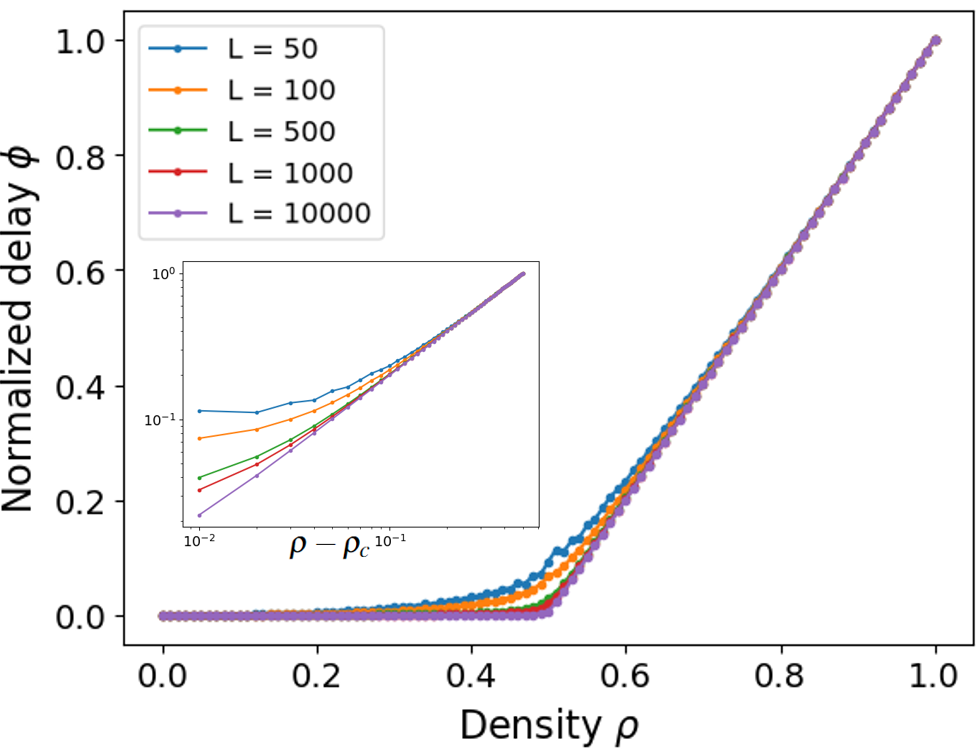} % Adjust the file name and path as needed
    \caption{We sample 1000 trials per density value for systems of various sizes $L$. (inset) log-log plot for various $\rho$ close to $\rho_c = 0.5$.}
    \label{fig:orderparam}
\end{figure}

\begin{figure}[htbp]
    \centering
    \includegraphics[width=0.5\textwidth]{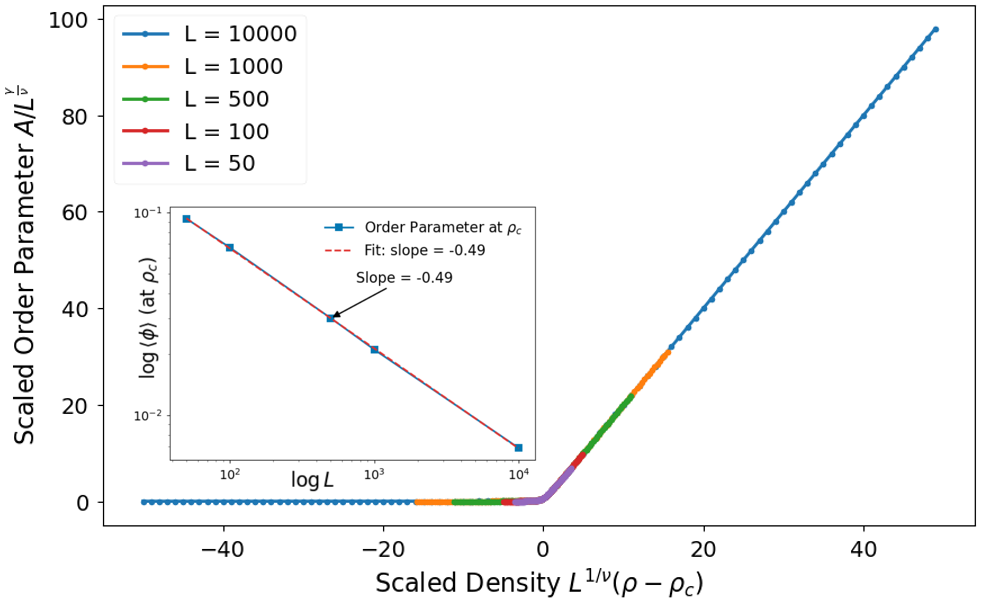} % Adjust the file name and path as needed
    \caption{Finite-size scaling collapse for the order parameter.  The scaling exponent $\nu = 2$ was obtained by visual inspection of data overlap. (inset) Exponent $\beta$ was obtained from normalised $\langle A \rangle$ at $\rho_c = 0.5$ for various system sizes $L$. The slope of the line gives us another way to obtain $\beta/\nu$. Here $\beta/\nu = 0.49$, which is close to our value of 1/2}
    \label{fig:orderparamfinitesize}
\end{figure}

\noindent Figure \ref{fig:orderparam} shows the results for $\phi$ versus density $\rho$. There is a transition at or close to $\rho$ = 0.5; the transition becomes sharper as the system size increases. The data resemble the behavior of an order parameter for a continuous phase transition, with $\rho_c$ = 0.5. Above the transition, we characterize the order parameter growth from a log-log plot (inset of fig \ref{fig:orderparam}): $\phi\:\sim\:|\rho -\rho_c|^{\beta}$ with critical exponent $\beta$.
% L becomes N here..
\noindent As the system size increases, the shape of the plot converges to a limit, but only rather slowly.  Thus, we employ finite-size scaling to obtain the value of the exponent $\beta$ for  $L\rightarrow\infty$. Specifically, we use
\begin{equation}
    \frac{\phi}{L^{-\beta/\nu}} = g(L^{1/\nu}(\rho-\rho_c))
\end{equation}
and determine the values of $\beta$ and $\nu$ for which the data sets for various $L$ collapse to a universal function $g$.  

%\noindent Here $-\beta/\nu$ is associated with how the $\langle A \rangle$ varies with N at $\rho_c$ and \textit{g} is a universal function different system sizes collapse to. 
%%%%IMPORTANT1
Figure \ref{fig:orderparamfinitesize} shows that very good data collapse is achieved, with $\nu=2$ and $\beta = 1$.  A more careful analysis to find the best fit results in the values $\nu = 2.0(0)$ and $\beta$ = 1.0(0) where the digits in parentheses represent uncertain digits in our analysis (details are given in the Appendix). These results are consistent with our hypothesis that $\nu = 2$ and $\beta = 1$

\begin{figure}[htbp]
    \centering
    \subfigure{
        \includegraphics[width=0.45\textwidth]{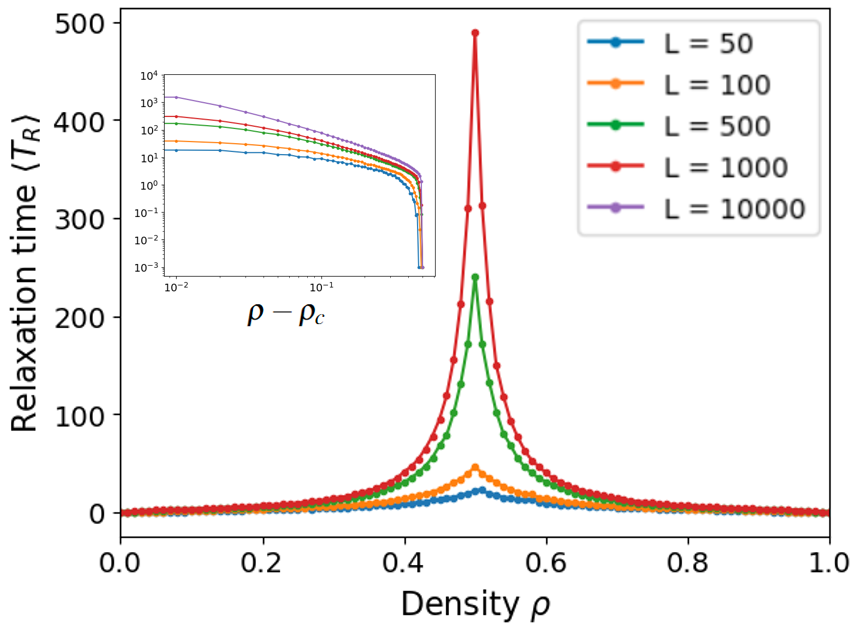}
        %\label{fig:relaxationtime}
    }
    \hfill % or use \quad or \hspace{1cm} for spacing
    \subfigure{
        \includegraphics[width=0.5\textwidth]{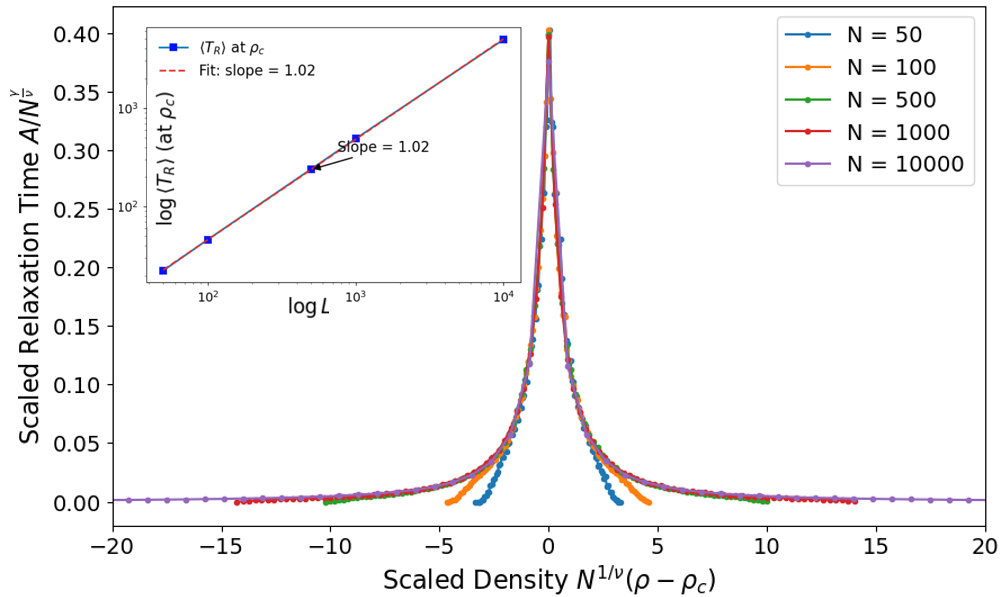}
        %\label{fig:relaxationfinitesize}
    }
    \caption{(a) Dependence of the $\langle T_R \rangle$ on $\rho$ for various system sizes $L$. The peak occurs at $\rho_c = 0.5$ and sharpens with increasing $L$. (inset) log-log plot for various $\rho$ close to $\rho_c = 0.5$.  (b) Finite-size scaling collapse for the relaxation time with exponents $\nu = 2$ and $\gamma = 2$. (inset) $\gamma/\nu$ was obtained from the slope of $log \langle T_R \rangle$ at $\rho_c = 0.5$ for various system sizes $log (L)$.}
    \label{fig:relaxationtime}
\end{figure}

\noindent Figure \ref{fig:relaxationtime}(a) shows the ensemble-averaged relaxation time  $\langle T_R\rangle $ as a function of $\rho$, which increases dramatically as $\rho$ approaches $\rho$ = 0.5 from either side, indicative of ``critical slowing down" typically observed in the vicinity of a critical point. From the inset, we observe that this slowing down is power law in nature:  $\langle T_R \rangle\:\sim\:|\rho -\rho_c|^{-\gamma}$.

As before, we obtain the critical exponent $\gamma$ for an infinite size system by applying finite-size scaling analysis.  Using:

\begin{equation}
    \frac{T_R}{L^{\gamma/\nu}} = f(L^{1/\nu}(\rho-\rho_c))
\end{equation} 

\noindent where $f$ is the same function for all data sets.  We already know that $T_R = \frac{L}{2}$ at $\rho_c$ (this is an exact result as $L\rightarrow\infty$), so that $\gamma/\nu = 1$.  Meanwhile, from data collapse we find $\nu = 2.0(2)$ and $\gamma = 2.0(4)$ (details described in Appendix).  Within uncertainty this is also in line with $\gamma = \nu =$ 2.  Thus,

\begin{equation}
    \langle T_R \rangle\:\sim\:|\rho -\rho_c|^{-\gamma}  \quad \text{with} \quad  \gamma = 2
\end{equation}
\noindent with the same exponent on either side of $\rho_c$. This yields the data collapse shown in Fig. \ref{fig:relaxationtime}(b). 
\subsection{Microscopic observables}
\noindent We examine the microscopic distributions from the space-time plot: cluster sizes $a_i$, lifetimes $\theta_i$, and elementary-jam lengths $m_i$ at the critical density $\rho_c = 0.5$. These distributions are obtained for various system sizes $L$, and we observe power-law regimes alongside finite-size effects.

For the cluster size distribution $p(a_i)$ we propose a finite-size scaling ansatz of the form:

\begin{equation}
    p(a, L) = a^{-\tau_a} f\left(\frac{a}{L^{D_f}}\right)
    \label{eq:scaling_ansatz}
\end{equation}

\noindent In Eq. \ref{eq:scaling_ansatz}, $\tau_a$ is the Fisher exponent and $D_f$ is the fractal dimension. From the data, we find $\tau_a = 1.5$ and $D_f = 1$. To address noise in the distribution, we also consider the survival function $s(a, L)$, which follows:

\begin{equation}
    s(a, L) = a^{-\alpha} f\left(\frac{a}{L^{D_f}}\right), \quad \alpha = \tau_a - 1
    \label{eq:survival_scaling}
\end{equation}

\begin{figure}[htbp]
    \centering
    \subfigure{\includegraphics[width=0.45\textwidth]{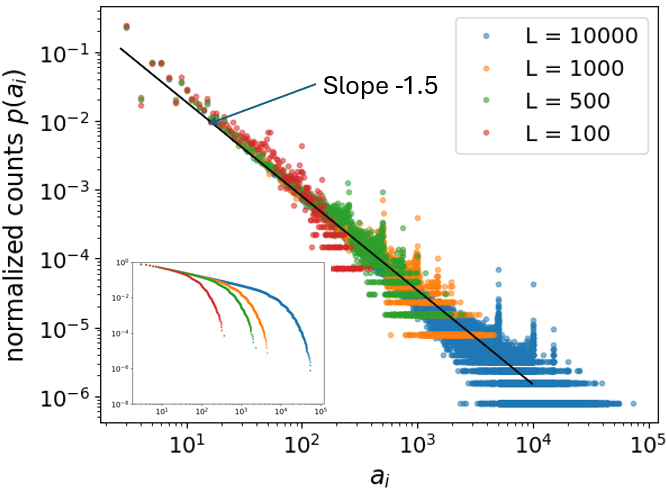}}
    \hfill
    \subfigure{\includegraphics[width=0.45\textwidth]{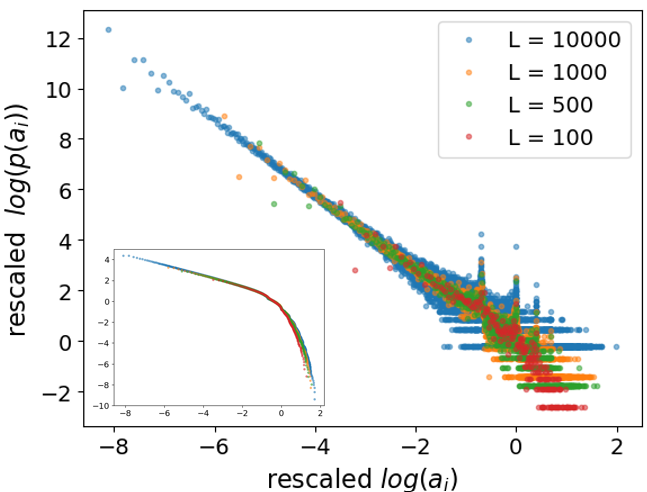}}
    \caption{(a) Cluster size distribution $p(a_i)$ for various system sizes $L$. (b) Finite-size scaling collapse of $p(a, L)$. Insets show the survival function $s(a, L)$, highlighting reduced noise and the appearance of wave-like features in the tail.}
    \label{fig:cluster_dist}
\end{figure}

\noindent Figure \ref{fig:cluster_dist}(a) shows the distribution $p(a_i)$ at $\rho = 0.5$ for different $L$. We observe a power-law decay, with distinct peaks at $L/2$, $3L/2$, etc., due to finite-size effects. The inset displays the survival function $s(a, L)$, where the power-law regime is clearer, and peaks appear as wave-like structures. The FSS collapse in Figure \ref{fig:cluster_dist}(b) confirms the scaling form in Eq. \ref{eq:scaling_ansatz}, with aligned peaks across system sizes.

\begin{figure}[htbp]
    \centering
    \subfigure{\includegraphics[width=0.45\textwidth]{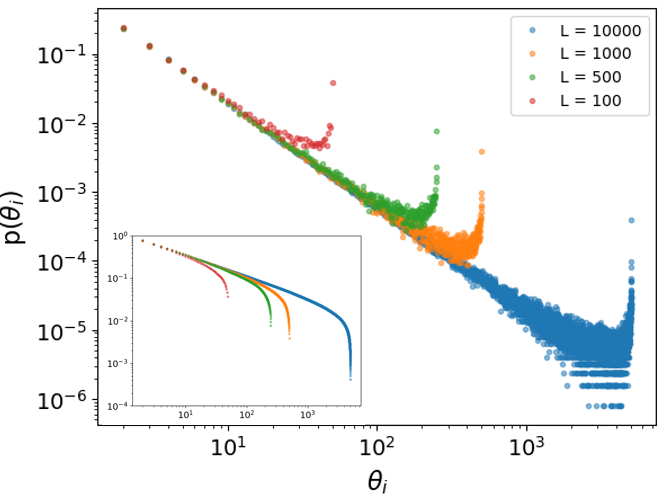}}
    \hfill
    \subfigure{\includegraphics[width=0.45\textwidth]{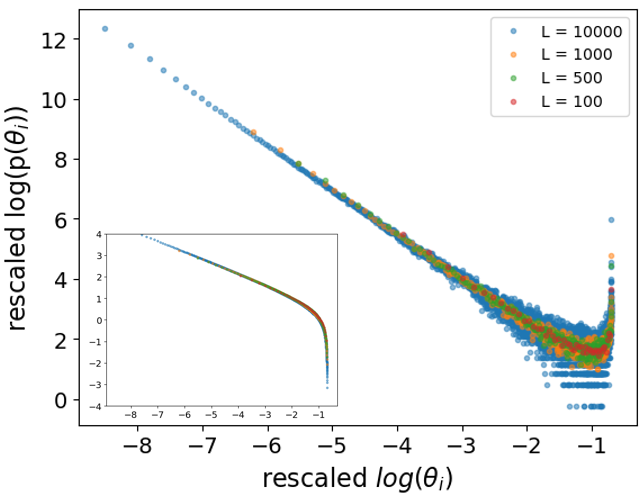}}
    \caption{(a) Lifetime distribution $p(\theta_i)$ for various system sizes $L$. (b) Finite-size scaling collapse of $p(\theta, L)$. Insets show the survival function $s(\theta, L)$, where the peak at $L/2$ is suppressed.}
    \label{fig:lifetime_dist}
\end{figure}

\noindent Figure \ref{fig:lifetime_dist}(a) presents the lifetime distribution $p(\theta_i)$. It follows a similar pattern to the cluster size distribution, with a power-law regime and a single prominent peak at $L/2$. The peak diminishes in the survival function $s(\theta, L)$, shown in the inset. We use the scaling form in Eq. \ref{eq:scaling_ansatz}, yielding $\tau_\theta = 1.5$ and $D_f = 1$, as seen in the FSS collapse in Figure \ref{fig:lifetime_dist}(b).

\begin{figure}[htbp]
    \centering
    \subfigure{\includegraphics[width=0.45\textwidth]{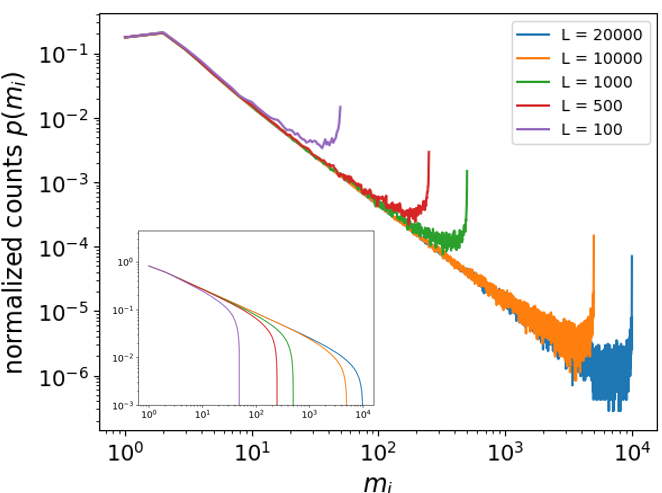}}
    \hfill
    \subfigure{\includegraphics[width=0.45\textwidth]{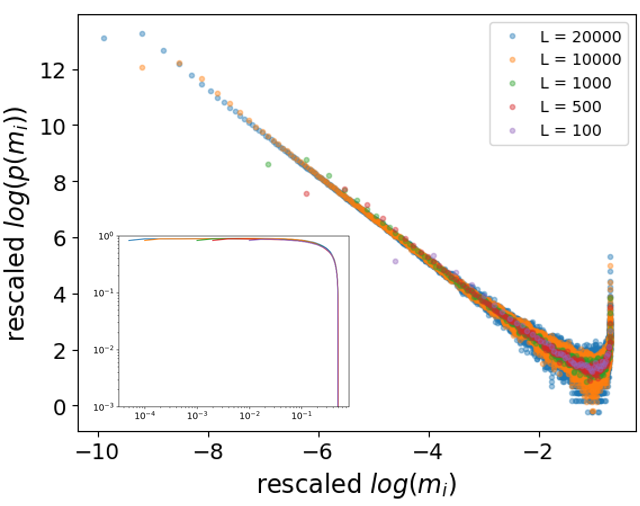}}
    \caption{(a) Elementary-jam length distribution $p(m_i)$ for various system sizes $L$. (b) Finite-size scaling collapse of $p(m, L)$. Insets show the survival function $s(m, L)$, where the peak at $L/2$ vanishes.}
    \label{fig:elementary_jam_dist}
\end{figure}

\noindent Figure \ref{fig:elementary_jam_dist} shows the distribution of elementary-jam lengths $p(m_i)$. It is similar to the lifetime distribution, with a power-law regime and a peak at $L/2$. In the survival function $s(m, L)$, shown in the inset, this peak is absent. Using Eq. \ref{eq:scaling_ansatz}, we obtain $\tau_m = 1.5$ and $D_f = 1$.

\begin{align*}
    p(a) &\sim a^{-\tau_a}; \quad \tau_a = 1.5(0) \\
    p(\theta) &\sim \theta^{-\tau_\theta}; \quad \tau_\theta = 1.5(0) \\
    p(m) &\sim m^{-\tau_m}; \quad \tau_m = 1.5(0)
\end{align*}

\noindent The consistent values of $\tau_a$, $\tau_\theta$, and $\tau_m$ across different observables suggest a universal scaling behavior, with $\tau \approx 1.5$. This scaling is consistent with previous work by Nagel and Paczuski \cite{nagel1995emergent}.

%----------------------------------
%\input{discussion}
%----------------------------------
\section{Summary and Discussion}

We have shown that ECA 184 exhibits features typical of a continuous phase transition. The macroscopic observables, such as the normalized delay $\phi$ and the relaxation time $T_R$, provide a clear picture of critical behavior. The order parameter $\phi$ increases continuously at the critical density $\rho_c = 0.5$, characterized by the scaling exponent $\beta = 1$. Meanwhile, the divergence of the relaxation time $T_R$ near $\rho_c$ with the exponent $\gamma = 2$ signals the phenomenon of critical slowing down, commonly seen in systems near criticality.

The finite-size scaling (FSS) analysis consistently yields the exponent $\nu = 2$ across independent methods. This agreement indicates that $\nu$ captures a meaningful physical correlation length in the system, diverging as $\chi \sim (\rho - \rho_c)^{-\nu}$. Notably, the equality $\gamma = \nu$ suggests a direct link between the system's response and its fluctuations, pointing towards a fluctuation-response relationship similar to those observed in equilibrium statistical mechanics.

The determination of the exponents $\gamma$, $\beta$, and $\nu$ through FSS methods parallels approaches used in the study of percolative transitions, such as in 2D percolation and directed percolation \cite{malthe2024percolation}. This analogy is supported by the observation that jammed vehicles in our model appear to percolate downwards along the time axis in the space-time plot. Following the perspective of Nagel and Paczuski in \cite{nagel1995emergent}, we can interpret this behavior as indicative of a percolative transition. In line with this, we propose the scaling form for the survival function of the lifetime distribution\cite{grassberger1979reggeon, nagel1995emergent}:

\begin{equation}
    s(\theta, \rho - \rho_c) \sim \theta^{-(\tau - 1)} f\left(\theta (\rho - \rho_c)^{\nu}\right).
    \label{eq:survival_function}
\end{equation}

This scaling form leads naturally to the relation $T_R \sim (\rho - \rho_c)^{-\nu}$, consistent with our finding that $\gamma = \nu$. Additionally, the relation $\beta = \nu(\tau - 1)$ aligns with the scaling laws expected for ECA 184, reflecting a coherent set of critical exponents.

The concept of elementary jams ($m_i$) plays a central role in our analysis, allowing us to express all relevant traffic observables in terms of these quantities. Unlike the macroscopic measures such as the relaxation time $T_R$ and the normalized total delay $\phi$, which serve as a response function and an order parameter respectively, the elementary jams are fundamentally space-time objects with no direct physical analog. Drawing from the language of directed percolation, we can interpret the jammed vehicles as ``active sites" that transition to ``inactive" states when they encounter voids. This perspective suggests that elementary jams act as propagators of activity, described by:

\begin{equation}
    m = p(\Delta x, \Delta t) \sim (\Delta x)^{-3/2} \delta(\Delta x - v_{jam} \Delta t),
\end{equation}
where $v_{jam} = -1$ and $\delta$ is the Dirac delta function. This propagator form captures the transmission of the ``jammed state" from one site to another over a space-time separation $(\Delta x, \Delta t)$.

The understanding of elementary jams as propagators provides a potential framework for analyzing similar models, such as the Biham-Middleton-Levine (BML) model and the Nagel-Schreckenberg (NS) model, where a comprehensive field-theoretic description remains an open question. Given the rational exponents observed, it is plausible to speculate that ECA 184 may correspond to a non-interacting field theory. In this context, ECA 184 could be viewed as an ``ideal gas" for traffic models, offering a baseline for exploring more complex, interacting traffic models.

%----------------------------------
%\input{conclusion}
%----------------------------------
\section{Conclusion}

We conclude that the transient jamming in our model may be described as a phase transition with a critical parameter value characterized by a complete set of exponents $\beta = 1$, $\gamma = 2$, $\nu = 2$ and $\tau = 3/2$. The simplicity of ECA184 allows us to develop efficient methods to calculate these space-time quantities and may provide deeper connections to an underlying field theory. Future work could explore the application of these ideas to more complex scenarios such as the Nagel-Schreckenberg model or road grids such as the BML automaton. Ultimately, these may be used to analyse real-world traffic networks and inform strategies for managing congestion.
%----------------------------------
\begin{acknowledgments}
This research was funded by NSF Award \#2311159. 
\end{acknowledgments}
\noindent\rule{\linewidth}{0.4pt}

\appendix

%_____Making appendix headings uppercase
\renewcommand{\appendixname}{\MakeUppercase{Appendix}} % Make the word 'APPENDIX' uppercase
\renewcommand{\thesection}{\Alph{section}} % Keep using letters (A, B, C...)
%______________

%\input{appendix}
%----------------------------------
\section{Exponent Uncertainty Analysis}

In the main text, we determined the critical exponents \(\nu\), \(\beta\), \(\gamma\), and \(\tau\) using finite-size scaling (FSS) analysis. To validate these values, we conducted a detailed examination of the uncertainty by varying the exponents slightly around their best-fit values and observing the quality of data collapse. Here, we present the analysis for the order parameter and lifetime distribution, which allowed us to visually obtain the best-fit values and estimate the uncertainty.

\begin{figure}[htbp]
    \centering
    \includegraphics[width=0.95\columnwidth]{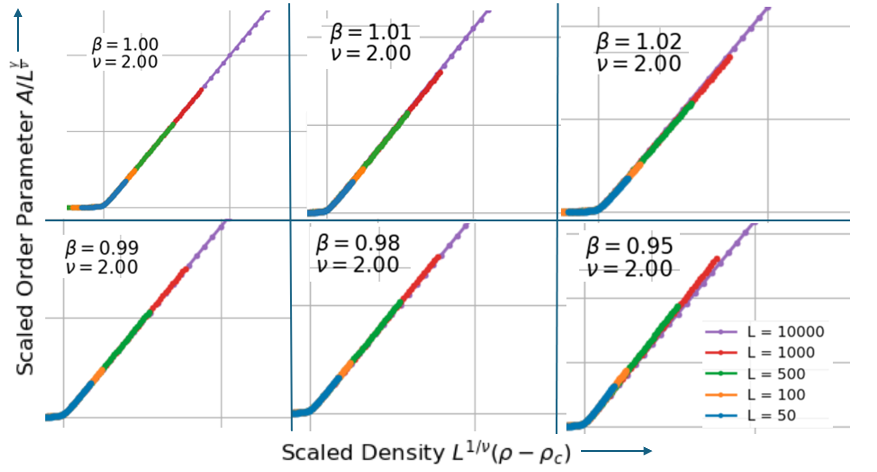}
    \caption{Finite-size scaling collapse for the order parameter \(\phi\) with varying \(\beta\). We fix \(\nu = 2\) and show the collapse for \(\beta = 1.01\), \(1.02\), \(0.99\), \(0.98\), and \(0.95\). The best fit with \(\beta = 1.00\) is highlighted. Deviations from the best-fit value result in poorer data collapse, indicating the range of uncertainty.}
    \label{fig:orderparam_uncertainty}
\end{figure}

Figure \ref{fig:orderparam_uncertainty} shows the FSS collapse for the order parameter \(\phi\), where we fixed \(\nu = 2\) and varied \(\beta\) as \(1.01\), \(1.02\), \(0.99\), \(0.98\), and \(0.95\). The best-fit value of \(\beta = 1.00\) is shown for reference. As \(\beta\) deviates from the best fit, the quality of the data collapse visibly worsens, particularly for \(\beta = 0.95\), where the deviation is most pronounced. This approach allows us to visually identify the range of \(\beta\) values that provide acceptable collapse, leading to an uncertainty estimate of \(\beta = 1.00 \pm 0.01\).

A similar analysis was performed for the other exponents. For \(\gamma\), we varied its value while keeping \(\nu = 2\) fixed, observing the collapse for relaxation time \(T_R\). Additionally, we varied \(\nu\) around the best-fit \(\beta = 1\) to obtain consistent results (not shown here).

\begin{figure}[htbp]
    \centering
    \includegraphics[width=0.9\columnwidth]{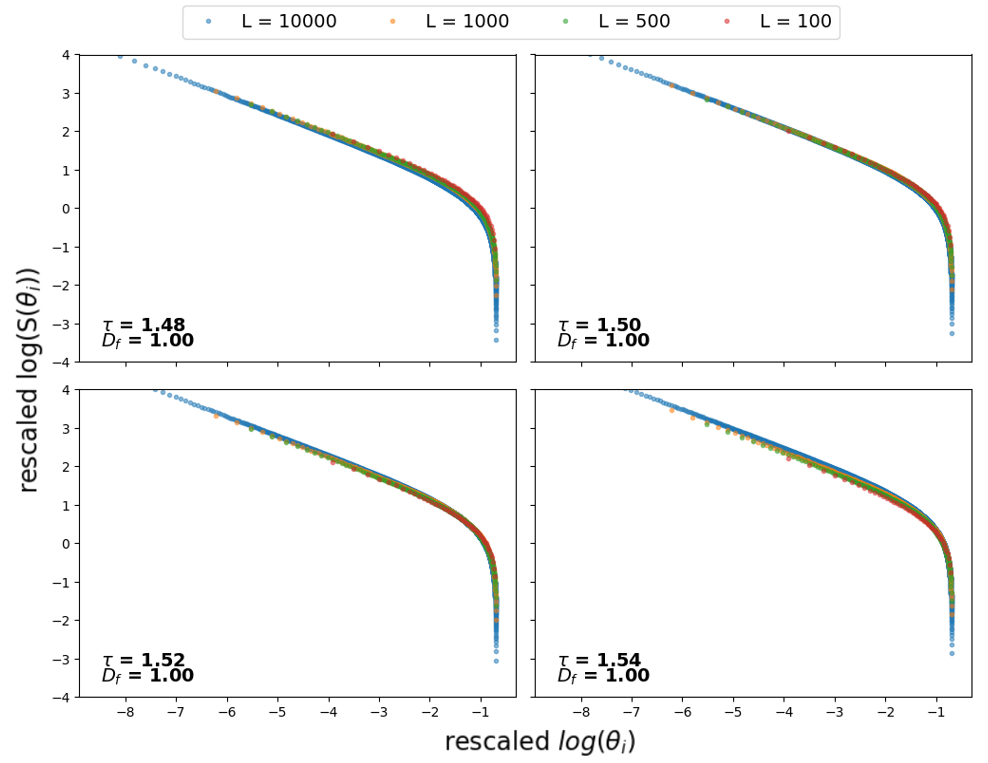}
    \caption{Finite-size scaling collapse for the lifetime distribution \(\theta\), analyzing the exponent \(\tau\). We varied \(\tau\) around the best-fit value of \(\tau = 1.5\), showing deviations for \(\tau = 1.48\) and \(\tau = 1.52\), which result in poorer data collapse. This confirms \(\tau = 1.5\) as the optimal estimate.}
    \label{fig:lifetime_uncertainty}
\end{figure}

To illustrate the analysis for microscopic observables, we focus on the lifetime distribution \(\theta\) in figure \ref{fig:lifetime_uncertainty}. The FSS collapse for \(\theta\) was analyzed by varying the exponent \(\tau\) around the best-fit value of \(\tau = 1.5\). As with the macroscopic case, deviations from the best fit (\(\tau = 1.48\) and \(\tau = 1.52\)) resulted in a noticeable reduction in the quality of the collapse, supporting the choice of \(\tau = 1.5\) as the most accurate estimate.

\section{Planar tree algorithm: Efficiently obtaining $m_j$}

\subsection{Runtime comparision}\label{subsec:Algo_time}
\begin{figure}[h! tbp]
    \centering
    \includegraphics[width=0.95\columnwidth]{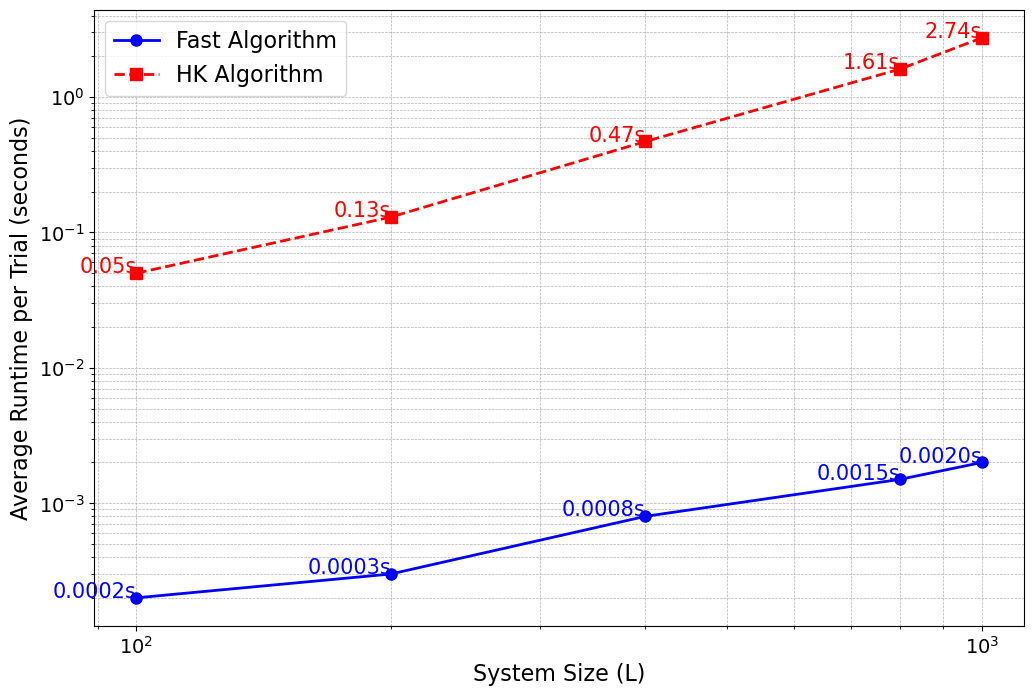} % Adjust the file name and path as needed
    \caption{Runtime of our tree algorithm and the HK algorithm {\it vs.} system size, averaging over a 1000 trials per system size. The runtime increases by a factor of 50 for the HK algorithm but only by a factor of 10 for the tree algorithm, which has $\sim O(L)$ runtime dependence.}
    \label{fig:runtime}
\end{figure}

The measurement of 2D connected clusters are commonly obtained by first generating the 2D space or space-time realization, and then applying the Hoshen-Kopelman (HK) algorithm \cite{hoshen1976percolation}, which allows one to obtain the cluster size distribution for arbitrary cluster shapes and is order $O(L^2)$ in runtime.  However, ECA184 generates cluster shapes that are quite limited in complexity, which we exploit to greatly increase the efficiency of determining the clusters and their properties.  Specifically, our algorithm determines the elementary jams directly from the initial conditions, and consequently the cluster properties $\left( a_i, \theta_i \right)$. See Figure \ref{fig:runtime} for a run-time comparison.

%\subsection{Algorithm details and Example}\label{subsec:Algo_details}

In what follows, we briefly outline our algorithm to obtain the distribution of elementary jam sizes $m_j$ from the initial condition.  

\subsection{Transforming Initial Conditions into Special Random walks}

\noindent Our algorithm begins by transforming the initial configuration of our system into a Dyck path, a special type of random walk. Unlike standard random walks, a Dyck path is constrained to stay non-negative. \cite{deutsch1999dyck}

\noindent We can illustrate this transformation step-by-step:

\begin{algorithm}[H]
\caption{Transformation of Initial Condition into Dyck Path}
\begin{algorithmic}[1]
\State \textbf{Input}: Initial condition \([1, 1, 0, 0, 0, 1]\)
\State \textbf{Step 1:} Replace each \(0\) with \(-1\) (step down) and keep \(1\) as \(+1\)(step up)
\State \hspace{2em} \(\rightarrow\) Transformed sequence: \([+1, +1, -1, -1, -1, +1]\)
\State \textbf{Step 2:} Reverse the sequence
\State \hspace{2em} \(\rightarrow\) Reversed sequence: \([+1, -1, -1, -1, +1, +1]\)
\State \textbf{Step 3:} Construct a random walk graph using the cumulative sum (Fig. \ref{fig:brownian_permuted}a)
\State \textbf{Step 4:} Identify the lowest point of the walk (marked as a red dot) and cyclically rotate the graph so this point is at the starting index (Fig. \ref{fig:brownian_permuted}b)
\State \textbf{Output}: Resulting Dyck path
\end{algorithmic}
\end{algorithm}

\begin{figure}[htbp]
    \centering
    \includegraphics[width=0.9\columnwidth]{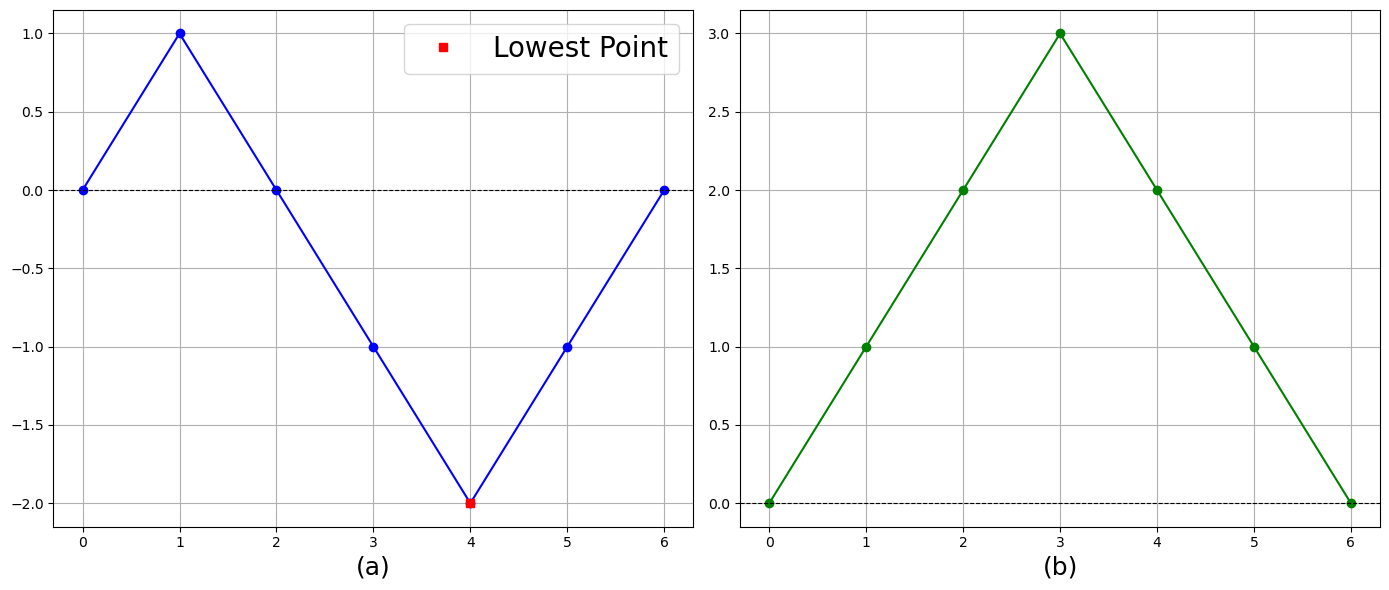}
    \caption{Transformation of the initial condition \([0, 0, 0, 1, 1, 0, 1, 0, 1, 1]\) into a Dyck path. (a) The cumulative sum representing the random walk. (b) The corresponding Dyck path after cyclic permutation.}
    \label{fig:brownian_permuted}
\end{figure}

\subsection{Tree-based Algorithm for Calculating Elementary Jam Lengths}

 The final step in our analysis is to extract the elementary jams $m_i$ from the Dyck path. These elementary jams are defined in section \ref{Elementary_jams} of the main text, and they can be determined using "first crossing times" of the Dyck path, a concept we now define.

\begin{figure}[htbp]
    \centering
    \includegraphics[width=0.9\columnwidth]{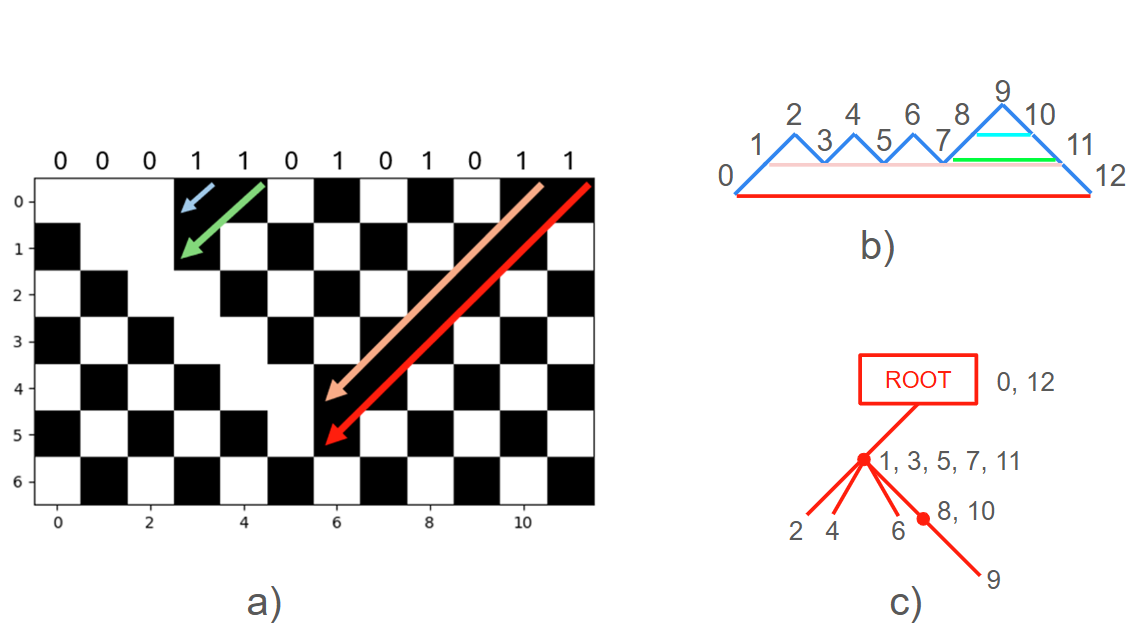}
    \caption{(a) Spacetime plot of ECA rule 184 showing elementary jams as colored diagonal arrows. (b) Dyck path representation with numbered steps highlighting the first crossing times, which determine the jam lengths. (c) Tree graph construction used to efficiently calculate the jam lengths \(m_i\) based on the first crossing times.}
    \label{fig:example_mi}
\end{figure}

\noindent Consider a given height \(h\) along the Dyck path. The first crossing time \(t_i\) at height \(h\) is the difference between the first time the path exceeds height \(h\) and the first subsequent time it falls below this height. Unlike a typical first return time, the first crossing time accounts for the path both reaching and crossing below the specified height. For example, in Figure~\ref{fig:example_mi}(b) index $1$ returns to its height at indices $3$ and $5$ and $7$ but crosses for the first time at index $9$. Index $7$ crosses for the first time at $9$ as well. The elementary jam length \(m_i\) can then be calculated using $m_i = \frac{t_i}{2}$

\noindent To efficiently compute \(t_i\), we use the following algorithm that generates a tree-graph from the Dyck path as shown in Figure~ \ref{fig:example_mi}c.

\begin{algorithm}[H]
\caption{Tree Graph Algorithm for Calculating Elementary Jam Lengths}
\begin{algorithmic}[1]
\State \textbf{Initialize}: Set the ROOT node at index 0 of the Dyck path.
\While{not at the end of the Dyck path}
    \If{current step is +1 (upward)}
        \State Create a child node for the current node.
    \ElsIf{current step is -1(downward)}
        \State Return to the parent node.
    \EndIf
\EndWhile
\State \textbf{Calculate first crossing times}: For each node, set \(t_i\) as the difference between the final index of the node and the indices corresponding to +1s which have at least another +1 as their neighbour.
\If{ROOT node starts with a jam}
    \State Set \(t_i\) as the difference between the first and last indices of the ROOT node.
\EndIf
\end{algorithmic}
\end{algorithm}

\noindent This tree-based approach allows us to quickly identify and measure the elementary jams, significantly speeding up the analysis compared to traditional methods. A comparison of the algorithm’s efficiency is provided in the first section of this Appendix. 

%----------------------------------
\onecolumngrid

\renewcommand{\bibsection}{\begin{center}\rule{0.5\textwidth}{0.4pt}\end{center}}
%% Bibliography section
%\bibliographystyle{unsrt}
%\bibliography{references}
% E. References Cited
%\newpage\setcounter{page}{1}
\renewcommand\refname{References Cited}
\bibliography{references}
\bibliographystyle{unsrtnat} 
\end{document}